 \documentclass[final,3p,times,twocolumn]{elsarticle}
 \usepackage{graphics}
\usepackage{amssymb}
\journal{Solid State Communications}

\begin{document}

\begin{frontmatter}

%% Title, authors and addresses

%% use the tnoteref command within \title for footnotes;
%% use the tnotetext command for theassociated footnote;
%% use the fnref command within \author or \address for footnotes;
%% use the fntext command for theassociated footnote;
%% use the corref command within \author for corresponding author footnotes;
%% use the cortext command for theassociated footnote;
%% use the ead command for the email address,
%% and the form \ead[url] for the home page:
%% \title{Title\tnoteref{label1}}
%% \tnotetext[label1]{}
%% \author{Name\corref{cor1}\fnref{label2}}
%% \ead{email address}
%% \ead[url]{home page}
%% \fntext[label2]{}
%% \cortext[cor1]{}
%% \address{Address\fnref{label3}}
%% \fntext[label3]{}

\title{Bandgaps and band bowing 
in semiconductor alloys}

%% use optional labels to link authors explicitly to addresses:
%% \author[label1,label2]{}
%% \address[label1]{}
%% \address[label2]{}

\author[1,2]{Titus Sandu}

\address[1]{International Centre of Biodynamics, Intrarea Portocalelor 
 Street, Nr.1B, Postal Code 060101, District 6, Bucharest, Romania}
\ead{tsandu@biodyn.ro}
\address[2]{D\'{e}partement de Chimie, Universit\'{e} de Montr\'{e}al, 
C.P. 6128, succursale Centre-ville, Montr\'{e}al, Qu\'{e}bec H3C 3J7, Canada}

\author[2]{Radu I. Iftimie}
\ead{radu.ion.iftimie@umontreal.ca}

\begin{abstract}
%% Text of abstract
The bandgap and band bowing parameter of semiconductor alloys are calculated with a fast and realistic approach. 
The method is a dielectric scaling approximation that is based on a scissor approximation. It adds an energy shift to the bandgap 
provided by the local density approximation (LDA) of the 
density functional theory (DFT).
The energy shift consists of a material-independent constant weighted by the inverse of the high-frequency 
dielectric constant. The salient feature of the approach is the fast calculation of 
the dielectric constant of alloys via the Green function (GF) of the TB-LMTO (tight-binding linear muffin-tin orbitals) in the atomic sphere approximation (ASA). 
When it is applied to 
highly mismatched semiconductor alloys (HMAs) like Zn Te$_x$ Se$_{1-x}$, this method provides a band bowing parameter that is 
different from the band bowing parameter calculated with the LDA due to the bowing exhibited also by the high-frequency 
dielectric constant.

\end{abstract}

\begin{keyword}
%% keywords here, in the form: keyword \sep keyword
A. disordered systems \sep A. semiconductors \sep D. electronic band structure \sep D. dielectric response
%% PACS codes here, in the form: \PACS code \sep code
\PACS 71.10.-w \sep  71.15.Ap \sep  71.55.Gs     
%% MSC codes here, in the form: \MSC code \sep code
%% or \MSC[2008] code \sep code (2000 is the default)

\end{keyword}

\end{frontmatter}

%% \linenumbers

%% main text
%%\section{}
%%\label{}

%% main text
\section{Introduction}
Semiconductor alloying is a very important way of bandgap engineering 
with applications 
in tailoring the materials properties like the electronic stucture or optical 
behaviour. The bandgap of semiconductor alloys can be usually described by a quadratic polynomial in 
concentration $x$ with the quadratic coefficient being called the bowing coefficient. 
Accordingly, the bandgap $E_g$ of an alloy AB$_{x}$C$_{1-x}$ is described by

\begin{equation}
\label{eq27}
E_{g}(x) = xE_{g}(AB) + (1-x)E_{g}(AC) - bx(1-x),
\end{equation}   

\noindent
where $E_{g}(AB)$ and $E_{g}(AC)$ are the bandgaps of the binary constituents $AB$ and $AC$. 
The bandgap bowing coefficient $b$ describes the deviation from linearity and is close to a 
constant for most alloys.

Highly mismatched semiconductor alloys (HMAs) exhibit a large bowing parameter. HMAs are present in 
III-V systems like  Ga N$_{x}$ As$_{1-x}$ \cite{Shan99}, Ga Bi$_{x}$ As$_{1-x}$, and Ga Sb$_{x}$ As$_{1-x}$ \cite{Alberi07}; 
II-VI compounds like Zn S$_x$ Te$_{1-x}$ \cite{Walukiewicz00},  Zn Te$_x$ Se$_{1-x}$\cite{Wu03}, and Zn O$_x$ Te$_{1-x}$ \cite{Yu03}; 
and IV-IV alloys like Sn$_x$ Ge$_{1-x}$ \cite{Alberi08}. 
Lattice relaxation, localization, charge transfer, and huge bandgap bowing are encountered in HMA systems due to large 
differences in atomic sizes and orbital energies, and large lattice mismatch.

The bandgap bowing of HMAs cannot be described well by simple models like virtual crystal approximation, at least 
in the mid-alloy regime of concentrations \cite{Tit09}. 
Other models like the band anticrossing (BAC) model \cite{Shan99}, the valence-band anticrossing (VBAC) models \cite{Alberi07} or 
combinations of them \cite{Wu03,Sandu05a} are able to explain the strong bandgap change with concentration $x$. 
These models are applied in the dilute limit of concentrations, in which one constituent is considered as 
an impurity with composition-dependent coupling to the conduction band minimum or valence band maximum. However, the BAC 
model and its extensions completely neglect cluster states induced by impurities \cite{Kent01}. Supercell 
methods overcome this shortcoming but the price is large unit cells \cite{Lindsay04}. More efficient approaches 
are based on special quasirandom structures (SQSs) having smaller supercells in which the most relevant atom-atom 
correlation functions are similar to those of random alloys  \cite{Zunger90}.

Standard first-principles band theory is based on the Kohn-Sham (KS) scheme \cite{KohnSham65} of 
the density functional theory (DFT) \cite{HohenbergKohn64}, which is, in principle, exact. 
The DFT is a total-energy method for calculating the electronic ground state. 
The local density approximation (LDA) \cite{HohenbergKohn64} and the generalized gradient approximation 
(GGA) \cite{Perdew96} provide reasonably accurate results for ground-state properties 
like crystal structure, pressure-induced transformations, phonon spectra, magnetism, etc. 
Usually the LDA leads to some overbindings. For example, in materials, where narrow bands determined by semi-core states 
(for instance, Zn 3d states in ZnSe and ZnTe) contribute to the bonding, the LDA 
predicts these localized semi-core states too high in energy. In addition, the LDA and GGA do not describe 
well the long-range part of correlations and have the 
so-called "bandgap problem". Even though the LDA often yields 
good band dispersions for the valence bands, the LDA bandgap energy in semiconductors and 
insulators is much smaller than the experimental bandgap. The theory 
that may solve that problem is the Hedin's GW approximation (GWA: $G=$ Green 
function and $W=$ screened Coulomb interaction) \cite{Hedin65,Aryasetiawan98,Schilfgaarde06}. 
The corrections in the GWA are provided by 
the self-energy that resembles the 
Hartree-Fock exchange term, which is a non-local operator. Fortunately the eigenvectors 
obtained with the Hartree-Fock approximation (HFA), the LDA, and the GWA  are often very close \cite{Hedin95,Puerto08}. 
Therefore, in many cases
one can obtain good results with the GWA by solving first the LDA problem, and then taking the expectation
value of the GWA self-energy over the LDA eigenvectors. 

The calculations in the GWA are based on the screening of the exchange interaction 
with the inverse frequency-dependent dielectric 
matrix as opposed to local exchange-correlation potentials in the LDA. The computational cost, however, is large due to the evaluation 
of dielectric matrices, their inversion, etc. Thus, the GWA is usually restricted to relatively small systems. 
One way to circumvent these computational hurdles is to use dielectric function models \cite{Gygi89,Bechstedt92} and 
dielectric models such as Thomas-Fermi 
screening in the screened-exchange LDA (SX-LDA) methods \cite{Bylander90,Seidl96}. 
Although these schemes are computationally ''cheaper'' than 
the GWA, only quite recently have SX-LDA methods been used to larger systems \cite{Furthmuller02,Lee06}. 
Moreover, despite the fact that these methods provide a much better description of the bandgaps than the LDA, it appears 
that it is rather difficult to define the screening constants of alloys.

Generally, in the HMA the bandgap bowing coefficient is 
composition dependent, reflecting the strong wave-function localization at the band edges 
(see \cite{Wu03} and the references therein). In such materials there is a strong 
interaction between the band edge states 
and localized, impurity-like states in alloys. To be explicit, in semiconductor 
alloys like ZnTe$_x$Se$_{1-x}$ there is a strong interaction between the localized Te level and the valence band 
of the host in the Se-rich limit of concentrations. On the other hand, in the Te-rich limit of 
concentrations there is a strong coupling between localized Se level and the conduction band. 
Because the bowing coefficient is calculated as a variation, i.e., a difference that enables some 
cancellations, the LDA would seem appropriate, at least in a zero-order approximation.
In alloys like ZnTe$_x$Se$_{1-x}$, however, the use of the LDA  has two major caveats. The first is 
related to the underestimation of the bandgap and the height of the conduction band. 
The second comes from the overbinding of the cation d states that mix too much with 
p states of the valence band maximum and push these p states upward in energy. Thus, in HMA the 
LDA would rather fail to have realistic assessment of not only the bandgap but also the band bowing 
over the whole range of alloy composition $x$. Below, we present a fast and simple method that 
estimates the bandgap and band bowing coefficient with an accuracy approaching that of 
more demanding methods like the GWA and SX-LDA. The concrete example of ZnTe$_x$Se$_{1-x}$ 
is analyzed and discussed.

\section{Method}

There is still a simpler method to calculate the semiconductor bandgaps. It is based on the scissor operator \cite{Aryasetiawan98} and 
is basically a dielectric scaling approximation (DSA) for correcting the LDA bandgaps  \cite{Fiorentini95}. Thus for a wide 
variety of materials the main correction to the LDA bandgap is \cite{Fiorentini95}

\begin{equation}
\label{eqband}
\Delta \simeq \frac{\alpha}{\varepsilon _\infty },
\end{equation}
where $\alpha=9.1$ eV is a universal constant and $\varepsilon _\infty $ is the high-frequency dielectric constant of the material. 
The method is applied not only to usual zinc blende or wurtzite types but also to other structures like technologically 
interesting high-dielectric materials \cite{Kersch08}.     
Beside the technical arguments presented in the original paper \cite{Fiorentini95} there are other arguments that ensure the success 
of this method. First, there is the perturbative argument of the GWA with respect the LDA \cite{Hedin95}: the GWA corrections are basically the expectation 
value of the GW self-energy over the LDA eigenvectors. Second, the difference between quasiparticle energies and the LDA eigenvalues is 
essentially due to the non-local nature of the effective potential \cite{Gruning06}, which can be expressed with a scissor-like operator \cite{Fiorentini95}. 
There is 
also a more physical picture provided by Harrison \cite{Harrison85}. The exact non-local exchange operator of the Hartree-Fock theory has a long-range 
part which is unscreened, 
and hence overestimated. In the LDA the exchange operator is replaced by a local exchange-correlation operator 
which neglects the long-range part of the exchange altogether. In semiconductors the long-range part of the 
exchange is screened by the high-frequency dielectric constant. The LDA uses a fixed one-electron potential for all bands including the 
unoccupied bands. Thus the LDA neglects the electrostatic energy $U$ needed for the separation of the electron and 
hole in the excitation that provides the bandgap. Harrison noted that the excitation process is accompanied by 
a dielectric relaxation around the electron and the hole such that the correction to the bandgap is a rigid 
shift for the entire conduction band of order $U/\varepsilon _\infty$. 
Finally, a quite similar approach has been reported recently. It starts 
from the optimized effective potential method with a free parameter, whose value is established by the separation 
between the short-ranged and long-ranged exchange in the screened hybrid functional \cite{Tran09}.

In the present work we calculate the bandgap and band bowing 
coefficient in semiconductor alloys beyond the LDA by calculating the high-frequency dielectric constant. 
The equilibrium positions of atoms in the SQS supercells representing the alloy are determined with 
the Vienna \textit{ab initio} 
simulation package (VASP) \cite{Kresse96} with the projector augmented-wave method (PAW) \cite{Kresse99}.
Then the electronic structure is calculated with the tight-binding (TB) linear muffin-tin orbital method (LMTO) in its atomic sphere 
approximation (ASA) \cite{Andersen84,Andersen86,Turek97} in the LDA.
The TB-LMTO-ASA method is one of the most efficient, versatile, and also one of the most physically transparent methods. 
Other methods use a bigger number of basis functions and they are usually more difficult to interpret physically. 
Even though it uses the shape approximation (the Wigner-Seitz (WS) cells are replaced by overlapping WS spheres), the TB-LMTO-ASA 
method is quite successful in the description of closely packed solids, whereas empty spheres are used in the interstitial region to 
close pack open structures \cite{Glotzel80}. As a rule, the shape approximation does quite well for closely packed systems, in 
particular for energy bands, but not that well for total energy calculations. 
Finally the static polarization $\chi^0$ and the high-frequency dielectric constant $\varepsilon_\infty$ are calculated with the TB-LMTO-ASA. 
The procedure is laid out below. 

Let us then consider the static screening of a test charge 
in the random phase approximation \cite{Aryasetiawan98}. Consider a lattice of points (spheres), with the electron density 
in equilibrium. We wish to calculate the response (screening) in the 
electron charge $\delta q_j $ at site $j$, induced by the addition of a small 
external potential $\delta V_i $ at site $i$. The electron charge $\delta q_j $ 
is related to $\delta V_i $ by the non-interacting response 
function $\chi_{ij}^0 $:

\begin{equation}
\label{eq19}
\delta q_i^0 \equiv \chi_{ij}^0 \delta V_j. 
\end{equation}

\noindent
$\chi_{ij}^0 $ can be calculated in the first-order perturbation theory from 
the eigenvectors of a non-interacting one-electron Schr\"{o}dinger equation with the 
Adler-Wiser formalism \cite{Adler62,Wiser63} or directly via $G^0$, the Green function (GF) calculated 
from the one-electron Hamiltonian of the LDA. By linearizing the 
Dyson equation for the perturbed GF, i. e. $\delta G = G^0\delta VG \approx G^0\delta VG^0$, 
one obtains an explicit representation of $\chi_{ij}^0 $ in 
terms of $G^0$

\begin{equation}
\label{eq20}
\begin{array}{l}
 \delta q_i^0 = \frac{1}{\pi }\mbox{Im}\sum\limits_\textbf{k} {\oint {dz\delta G_{ii}(\textbf{k},z) } }  \\ 
 \quad = \frac{ 1}{2\pi i }\sum\limits_\textbf{k} {\oint {dzG_{ij}^0(\textbf{k},z) G_{ji}^0(\textbf{k},z) } 
\delta V_j^0 } \\ 
 \quad = \chi_{ij}^0 \delta V_j^0. \\ 
 \end{array}
\end{equation}

\noindent
In Eq. (\ref{eq20}) the $\textbf{k}$ summation is the summation in the Brillouin zone and 
the energy-integration contour encloses the occupied states. 
A large part of the computer resources is allocated to the calculation of $\chi^0$. 
The calculation of $\chi^0$ by the Adler-Wiser formula would require the utilization of large 
computer resources because it needs all the occupied orbitals at once.
In contrast, Eq. (\ref{eq20}) shows clearly that $\chi^0$ can be calculated piecewisely, and hence efficiently once 
the GF is obtained. 
Equation (\ref{eq20}) is the TB-LMTO-ASA counterpart of the static polarization function defined in real space \cite{Aryasetiawan98}. 
Similarly, the dielectric function is 

%\begin{widetext}
\begin{equation}
\label{eq21}
\varepsilon  = 1 -  V \chi^0 ,
\end{equation}
%\end{widetext}

\noindent
where $V$ is the unscreened Coulomb matrix. 
The macroscopic dielectric constant $\epsilon_\infty $ is calculated as \cite{Baroni86}

\begin{equation}
\label{eq28}
\varepsilon _\infty = \frac{1}{N_o }\sum\limits_\beta {\mathop 
{\mbox{lim }}\limits_{\textbf{k} \to 0} 1/[\varepsilon_{\beta ,\beta } \left( \textbf{k} \right)}]^{-1}. 
\end{equation}  

\noindent
In equation (\ref{eq28}) $N_o$ is the total number of orbitals. The indices $\beta = \left( { \textbf{R},l,m} \right)$ denote the 
lattice sites \textbf{R} and the angular-momentum quantum numbers $l$ 
and $m$. 
Nevertheless equation (\ref{eq28}) neglects "local field" effects and it represents the upper bound 
of $\epsilon_\infty $ when "local field" effects are considered \cite{Baroni86}.

The GF can be 
directly calculated within the LDA of the TB-LMTO-ASA by 
using the potential parameters,  $C$, $\Delta $, and $\gamma $ \cite{Andersen84,Andersen86,Turek97}

\begin{equation}
\label{eq23}
\begin{array}{ll}
G\left( z \right) =  \\
- \frac{1}{2}\frac{\ddot {P}^\alpha \left( z 
\right)}{\dot {P}^\alpha \left( z \right)} +  \sqrt {\dot {P}^\alpha \left( z 
\right)} \left[ {P^\alpha \left( z \right) - S^\alpha } \right]^{ - 1}\sqrt 
{\dot {P}^\alpha \left( z \right)} \\
\end{array}
\end{equation}

\noindent
where \textit{$\alpha $} is the representation, determined by the screening constants $\alpha 
_{Rl} $ and

\begin{equation}
\label{eq24}
P_{Rl}^\alpha \left( z \right) = \frac{z - C_{Rl} }{\Delta _{Rl} + \left( 
{\gamma _{Rl} - \alpha _{Rl} } \right)\left( {z - C_{Rl} } \right)}
\end{equation}

\noindent
is the potential function. $\dot {P}^\alpha $ and $\ddot {P}^\alpha $ are, respectively, 
the first  and the second derivatives of the potential function with respect to energy. $S^\alpha =S_{\textbf{R}lm,\textbf{R}'l'm'}^\alpha$ 
is the matrix of structure constants in the $\alpha $ 
representation. In the 
LMTO-ASA approach, the effective potential is entirely described by the potential parameters:
the band center $C$, bandwidth $\Delta $, and the distortion parameter $\gamma $. The potential function provides 
the boundary conditions for the radial Schr\"{o}dinger 
equation inside each WS sphere \cite{Andersen84,Andersen86,Skriver84}. The fact that both the charge density and the effective 
potential are described by a few atomic parameters makes the calculation of the response functions a very simple task.

\begin{figure}
\includegraphics [width=2.75in]  {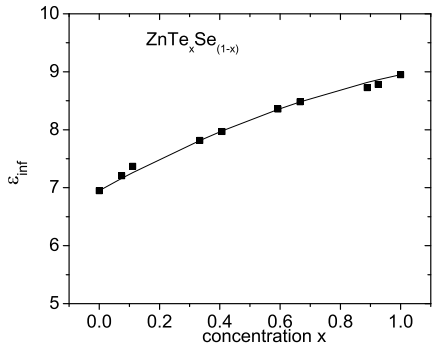}
\caption{\label{fig1} Macroscopic dielectric constant $\varepsilon _\infty$ estimated with the TB-LMTO-ASA. There is an estimated bowing of about $-0.89$} 
\end{figure}

\section{Results and Discussions}

For illustration we show 
the calculations of the bandgap corrections in the ZnTe$_x$Se$_{1-x}$ alloys. 
The random zinc blende alloys were modeled as supercells of 54 atoms (27 cations) with 
special quasirandom structures (SQSs) \cite{Zunger90}. The supercells have the lattice vectors 
$(3\textbf{a}_1,3\textbf{a}_2,3\textbf{a}_3)$ with $(\textbf{a}_1,
\textbf{a}_2,\textbf{a}_3)$ the unit cell vectors of the binary constituents. 
The SQS structures are constructed such that the most relevant atom-atom correlation 
functions of the SQS structure are similar to those of random 
alloys \cite{Zunger90}. We assumed that the alloy lattice constants are 
determined by the weighted average of the lattice constants of the 
constituents (Vegard's law) \cite{Vegard21}. In the ASA, the 
spherical average of the charge in each sphere makes the TB-LMTO approach less reliable and more 
cumbersome in calculating the equilibrium positions of the atoms in the SQS cells. For that reason the 
atomic positions inside the SQS cells were relaxed using VASP with the LDA exchange-correlation potential; at 
relaxed positions, the quantum-mechanical forces on each atom were less 
than $0.03$ eV/\AA. As we have already mentioned the atomic sphere approximation assumes that the whole 
space is filled with (overlapping) spheres and that the volume of the interstitial 
region vanishes. This is achieved by adding empty spheres \cite{Glotzel80}. 
The empty spheres must satisfy several criteria: the sum of the 
volume of all spheres should equal the volume of all space, the average overlap 
between the spheres should be minimal, and the radii of the spheres should be chosen 
such that the spheres overlap in the regions close to the local maxima of 
the electrostatic potential. The positions of the empty spheres are well known 
in the high-symmetry systems like zinc blende structures. In relaxed SQS 
cells the positions and the radii of the empty spheres were chosen to satisfy the 
above criteria starting from the initial positions of the non-relaxed 
(higher-symmetry) SQS cells. The Brillouin zone integration was performed on a 
$4 \times 4 \times 4$ mesh by a Monkhorst-Pack $k$-point sampling scheme \cite{Monkhorst76}.

In Fig. \ref{fig1} we plot the calculated dielectric constant $\epsilon_\infty$. 
 The calculated 
values of $\epsilon_\infty$ for the binary constituents are $10 - 20 \%$ larger than the 
experimental values and $5 - 15 \%$ larger than the time-dependent density functional theory 
(TD-DFT) calculations \cite{Kootstra00}. There is a bowing shown also by the macroscopic dielectric constant 
$\epsilon_\infty $, which is $b \approx -0.89$. Equation (\ref{eqband}) and the bowing exhibited by $\epsilon_\infty$ show us that the 
DFT-LDA alone cannot 
reproduce the bowing of the bandgap in ZnTe$_{x}$Se$_{1-x}$.

\begin{figure}
\includegraphics [width=2.75in] {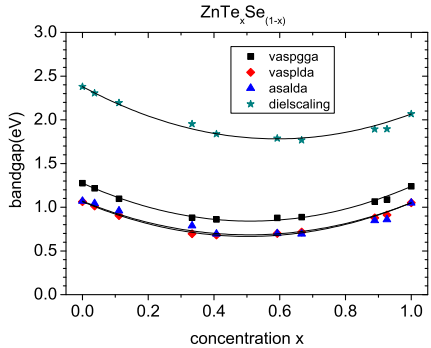}
\caption{\label{fig:2} The bandgaps of ZnTe$_x$Se$_{1-x}$ calculated with the LDA, the GGA, and Eq. 
(\ref{eqband}) at various compositions $x$ . The squares and the diamonds indicate the GGA-VASP and the LDA-VASP results, respectively; the 
triangles and the stars represent the values of the LDA-TB-LMTO-ASA and of the DSA, respectively. The continuous lines are the results 
of the fitting. No spin-orbit coupling was considered.
}
\end{figure}

Fig. \ref{fig:2} shows the 
bandgaps calculated with GGA-VASP and LDA-VASP, with the TB-LMTO-ASA in the LDA, and with the DSA defined by (\ref{eqband}). 
The TB-LMTO-ASA results were considered with combined corrections \cite{Skriver84}.
The bandgaps calculated with GGA-VASP are larger than the results of LDA-VASP over entire range of concentrations. 
The bandgaps of the binary constituents calculated with LDA-VASP are both 
1.06 eV. The LDA ASA gives a bandgap value of 1.07 eV for ZnSe and 1.05 eV 
for ZnTe. The ASA values are pretty close to the LMTO full potential values which are 
1.05 eV and 1.03 eV, respectively \cite{Schilfgaarde06}. Over the whole range of Te concentration $x$, 
the LDA-ASA results are in quite good agreement with the results obtained with PAW-VASP, which, in principle, 
is a full potential method with no shape approximation. 
Fig. \ref{fig:2} also shows the bandgaps calculated with the DSA, which exhibits much larger 
bandgaps than both the LDA and the GGA. The bandgaps of ZnSe and ZnTe calculated with the DSA are 2.38 eV and 2.07 eV, respectively, 
being comparable with the 
results of the GWA, which, in general, underestimates the bandgaps because it generates too much screening due to 
overestimation of $\chi^0$ \cite{Schilfgaarde06}. 
   
As we expected, the bandgaps 
show a bowing, i.e., the bandgap deviates from linear interpolation with respect to 
composition $x$. The bowing parameter $b$ is 1.64 eV for the VASP-GGA and 1.57 eV for the VASP-LDA. The LDA of the TB-LMTO-ASA produces a 
bowing of 1.50 eV, while the DSA estimates the bowing parameter to 1.71 eV. The bowing parameters predicted with methods that go beyond the LDA 
are larger. The fact that the LDA does not reproduce 
the entire bowing of the bandgap is rooted not only in the LDA's inability to predict the bandgap but also in the overbinding produced by the LDA.  
For ZnTe, ZnSe, and ZnTe$_{x}$Se$_{1-x}$ the LDA overbinding pushes the cation d levels closer to the top of the valence band. Thus the $d$ 
levels will mix too much with the 
p states at the maximum of the valence band; therefore additional errors in 
estimating the band bowing occur. On the other hand it is known that the GWA moves these cation $d$ states in the right direction downwards \cite{Schilfgaarde06}.

Old measurements of the bandgap bowing parameters $b$ with and without spin-orbit interaction are 
presented in \cite{Ebina74}. The authors estimated that the bandgap bowing is $b=1.266$ eV. They also estimated the 
spin-orbit splitting, which shows a concave bowing. Thus their estimated bandgap bowing without spin-orbit coupling is 
 $b=1.07$ eV. Other measurements estimate a bowing parameter of 1.62 eV (at 5 K) and 1.51 eV (at 300 K) \cite{Brasil91}. 
More recent experiments set the bowing of the bandgap between 1.25 eV and 1.5 eV at 300 K \cite{Wu03,Wu03b} and the bowing of the 
spin-orbit splitting at -0.33 eV \cite{Wu03}. Other measurements at low temperature, namely at 13 K, indicate a bowing 
parameter around 2.21 eV \cite{Lin08}.     

Several general features can be outlined from all experimental data and our calculations. More recent experiments 
reveal larger band bowing coefficients than the older ones. The bowing is larger at low temperatures. There is also 
a bowing of the spin-orbit splitting but it is negative. The bandgap bowing calculated with spin-orbit coupling 
is larger than the bandgap bowing without spin-orbit coupling by one third of the bowing parameter modulus of the spin-orbit splitting. 
Overall, our results for bandgap bowing are in rather good agreement with recent experimental results.

\section{Summary and Conclusions}
To summarize, we have presented a computational scheme that realistically estimates the bandgaps 
and band bowing parameters with applicability to 
semiconductors and their alloys. The scheme uses the static polarization function $\chi^0$ to calculate the high-frequency 
dielectric constant $\varepsilon _\infty$ in the GF-TB-LMTO formalism. 
The corrections to the LDA bandgaps are made with a dielectric scaling approximation (DSA), 
which estimates the correction to the LDA bandgap as a universal constant divided by the high-frequency 
dielectric constant. 

The approach inherits its speed from the fact that it works piecewisely 
to calculate $\chi^0$ via GF formalism and its efficiency from 
the TB-LMTO (the use of the muffin-tin orbitals and a minimal basis). 

The limitations of the method were already discussed in \cite{Fiorentini95}. The narrow-bandgap 
semiconductors are particularly difficult to describe with the DSA because the other corrections of the 
GWA to the the LDA bandgap become as important as the DSA. Also, often in these semiconductors there might appear 
a wrong ordering of the LDA levels around the bandgap with poor results in GWA too \cite{Schilfgaarde06}.

We applied the scheme to the ternary alloy ZnTe$_{x}$Se$_{1-x}$ by calculating the bandgap and band bowing in a 
supercell with 54 atoms that mimics the alloying of ZnTe$_{x}$Se$_{1-x}$. The ternary alloy ZnTe$_{x}$Se$_{1-x}$ 
exhibits large bandgap bowing. The bandgap bowing computed with the DSA is in good agreement with the experimental data. 
It was also shown that the bowing of the bandgap in ZnTe$_{x}$Se$_{1-x}$ cannot be explained by the LDA alone due to inherent 
bowing shown also by the dielectric constant. 

\section{Acknowledgments}
This work has been supported in part by the Romanian Ministry of Education and Research under the project ``Ideas'' No.120/2007.
The work was also partially supported by the Natural Science and Engineering Research Council of Canada (NSERC) grant and was made 
possible by the computational resources of the Reseau Qu\'{e}becois de Calcul de Haute Performance (RQCHP).

%\section*{References}
\bibliographystyle{elsarticle-num-names}

%\begin{thebibliography}{72}
%\bibliography{screening-effects}

\begin{thebibliography}{47}
\providecommand{\natexlab}[1]{#1}
\providecommand{\url}[1]{\texttt{#1}}
\providecommand{\urlprefix}{URL }
\expandafter\ifx\csname urlstyle\endcsname\relax
  \providecommand{\doi}[1]{doi:\discretionary{}{}{}#1}\else
  \providecommand{\doi}[1]{doi:\discretionary{}{}{}\begingroup
  \urlstyle{rm}\url{#1}\endgroup}\fi
\providecommand{\bibinfo}[2]{#2}

\bibitem[{Shan et~al.(1999)Shan, Walukiewicz, III, Haller, Geisz, Friedman,
  Olson, and Kurz}]{Shan99}
\bibinfo{author}{W.~Shan}, \bibinfo{author}{W.~Walukiewicz},
  \bibinfo{author}{J.~W.~A. III}, \bibinfo{author}{E.~E. Haller},
  \bibinfo{author}{J.~F. Geisz}, \bibinfo{author}{D.~J. Friedman},
  \bibinfo{author}{J.~M. Olson}, \bibinfo{author}{S.~R. Kurz},
  \bibinfo{journal}{Phys. Rev. Lett.} \bibinfo{volume}{82}
  (\bibinfo{year}{1999}) \bibinfo{pages}{1221}.

\bibitem[{Alberi et~al.(2007)Alberi, Wu, Walukiewicz, Yu, Dubon, Watkins, Wang,
  Liu, Cho, and Furdyna}]{Alberi07}
\bibinfo{author}{K.~Alberi}, \bibinfo{author}{J.~Wu},
  \bibinfo{author}{W.~Walukiewicz}, \bibinfo{author}{K.~M. Yu},
  \bibinfo{author}{O.~D. Dubon}, \bibinfo{author}{S.~P. Watkins},
  \bibinfo{author}{C.~X. Wang}, \bibinfo{author}{X.~Liu},
  \bibinfo{author}{Y.-J. Cho}, \bibinfo{author}{J.~Furdyna},
  \bibinfo{journal}{Phys. Rev. B} \bibinfo{volume}{75} (\bibinfo{year}{2007})
  \bibinfo{pages}{045203}.

\bibitem[{Walukiewicz et~al.(2000)Walukiewicz, Shan, Yu, III, Haller,
  Miotkowski, Seong, Alawadhi, and Ramdas}]{Walukiewicz00}
\bibinfo{author}{W.~Walukiewicz}, \bibinfo{author}{W.~Shan},
  \bibinfo{author}{K.~M. Yu}, \bibinfo{author}{J.~W.~A. III},
  \bibinfo{author}{E.~E. Haller}, \bibinfo{author}{I.~Miotkowski},
  \bibinfo{author}{M.~J. Seong}, \bibinfo{author}{H.~Alawadhi},
  \bibinfo{author}{A.~K. Ramdas}, \bibinfo{journal}{Phys. Rev. Lett.}
  \bibinfo{volume}{85} (\bibinfo{year}{2000}) \bibinfo{pages}{1552}.

\bibitem[{Wu et~al.(2003{\natexlab{a}})Wu, Walukiewicz, Yu, III, Haller,
  Miotkowski, Ramdas, Su, Sou, Perera, and Denlinger}]{Wu03}
\bibinfo{author}{J.~Wu}, \bibinfo{author}{W.~Walukiewicz},
  \bibinfo{author}{K.~M. Yu}, \bibinfo{author}{J.~W.~A. III},
  \bibinfo{author}{E.~E. Haller}, \bibinfo{author}{I.~Miotkowski},
  \bibinfo{author}{A.~K. Ramdas}, \bibinfo{author}{C.~H. Su},
  \bibinfo{author}{I.~K. Sou}, \bibinfo{author}{R.~C.~C. Perera},
  \bibinfo{author}{J.~D. Denlinger}, \bibinfo{journal}{Phys. Rev. B}
  \bibinfo{volume}{67} (\bibinfo{year}{2003}{\natexlab{a}})
  \bibinfo{pages}{035207}.

\bibitem[{Yu et~al.(2003)Yu, Walukiewicz, Wu, Shan, Beeman, Scarpulla, Dubon,
  and Becla}]{Yu03}
\bibinfo{author}{K.~M. Yu}, \bibinfo{author}{W.~Walukiewicz},
  \bibinfo{author}{J.~Wu}, \bibinfo{author}{W.~Shan}, \bibinfo{author}{J.~W.
  Beeman}, \bibinfo{author}{M.~A. Scarpulla}, \bibinfo{author}{O.~D. Dubon},
  \bibinfo{author}{P.~Becla}, \bibinfo{journal}{Phys. Rev. Lett.}
  \bibinfo{volume}{91} (\bibinfo{year}{2003}) \bibinfo{pages}{246403}.

\bibitem[{Alberi et~al.(2008)Alberi, Blacksberg, Nikzad, Yu, Dubon, and
  Walukiewicz}]{Alberi08}
\bibinfo{author}{K.~Alberi}, \bibinfo{author}{J.~Blacksberg},
  \bibinfo{author}{L.~D. B.~S. Nikzad}, \bibinfo{author}{K.~M. Yu},
  \bibinfo{author}{O.~D. Dubon}, \bibinfo{author}{W.~Walukiewicz},
  \bibinfo{journal}{Phys. Rev. B} \bibinfo{volume}{77} (\bibinfo{year}{2008})
  \bibinfo{pages}{073202}.

\bibitem[{Tit et~al.(2009)Tit, Obaidat, and Alawadhi}]{Tit09}
\bibinfo{author}{N.~Tit}, \bibinfo{author}{I.~M. Obaidat},
  \bibinfo{author}{H.~Alawadhi}, \bibinfo{journal}{J. Phys.: Condens. Matter}
  \bibinfo{volume}{21} (\bibinfo{year}{2009}) \bibinfo{pages}{075802}.

\bibitem[{Sandu and Kirk(2005)}]{Sandu05a}
\bibinfo{author}{T.~Sandu}, \bibinfo{author}{W.~P. Kirk},
  \bibinfo{journal}{Phys. Rev. B} \bibinfo{volume}{72} (\bibinfo{year}{2005})
  \bibinfo{pages}{073204}.

\bibitem[{Kent and Zunger(2001)}]{Kent01}
\bibinfo{author}{P.~R.~C. Kent}, \bibinfo{author}{A.~Zunger},
  \bibinfo{journal}{Phys. Rev. B} \bibinfo{volume}{64} (\bibinfo{year}{2001})
  \bibinfo{pages}{115208}.

\bibitem[{Lindsay and O'Reilly(2004)}]{Lindsay04}
\bibinfo{author}{A.~Lindsay}, \bibinfo{author}{E.~P. O'Reilly},
  \bibinfo{journal}{Phys. Rev. Lett.} \bibinfo{volume}{93}
  (\bibinfo{year}{2004}) \bibinfo{pages}{196402}.

\bibitem[{Zunger et~al.(1990)Zunger, Wei, Ferreira, and Bernard}]{Zunger90}
\bibinfo{author}{A.~Zunger}, \bibinfo{author}{S.~W. Wei},
  \bibinfo{author}{L.~G. Ferreira}, \bibinfo{author}{J.~E. Bernard},
  \bibinfo{journal}{Phys. Rev. Lett.} \bibinfo{volume}{65}
  (\bibinfo{year}{1990}) \bibinfo{pages}{353}.

\bibitem[{Kohn and Sham(1965)}]{KohnSham65}
\bibinfo{author}{W.~Kohn}, \bibinfo{author}{L.~J. Sham},
  \bibinfo{journal}{Phys. Rev.} \bibinfo{volume}{140} (\bibinfo{year}{1965})
  \bibinfo{pages}{A1133}.

\bibitem[{Hohenberg and Kohn(1964)}]{HohenbergKohn64}
\bibinfo{author}{P.~Hohenberg}, \bibinfo{author}{W.~Kohn},
  \bibinfo{journal}{Phys. Rev.} \bibinfo{volume}{136} (\bibinfo{year}{1964})
  \bibinfo{pages}{B864}.

\bibitem[{Perdew et~al.(1996)Perdew, Burke, and Ernzerhof}]{Perdew96}
\bibinfo{author}{J.~P. Perdew}, \bibinfo{author}{K.~Burke},
  \bibinfo{author}{M.~Ernzerhof}, \bibinfo{journal}{Phys. Rev. Lett.}
  \bibinfo{volume}{77} (\bibinfo{year}{1996}) \bibinfo{pages}{3865}.

\bibitem[{Hedin(1965)}]{Hedin65}
\bibinfo{author}{L.~Hedin}, \bibinfo{journal}{Phys. Rev.} \bibinfo{volume}{139}
  (\bibinfo{year}{1965}) \bibinfo{pages}{A796}.

\bibitem[{Aryasetiawan and Gunnarsson(1998)}]{Aryasetiawan98}
\bibinfo{author}{F.~Aryasetiawan}, \bibinfo{author}{O.~Gunnarsson},
  \bibinfo{journal}{Rep. Prog. Phys.} \bibinfo{volume}{61}
  (\bibinfo{year}{1998}) \bibinfo{pages}{237}.

\bibitem[{van Schilfgaarde et~al.(2006)van Schilfgaarde, Kotani, and
  Faleev}]{Schilfgaarde06}
\bibinfo{author}{M.~van Schilfgaarde}, \bibinfo{author}{T.~Kotani},
  \bibinfo{author}{S.~V. Faleev}, \bibinfo{journal}{Phys. Rev. B}
  \bibinfo{volume}{74} (\bibinfo{year}{2006}) \bibinfo{pages}{245125}.

\bibitem[{Hedin(1995)}]{Hedin95}
\bibinfo{author}{L.~Hedin}, \bibinfo{journal}{Int. J. Quantum Chem.}
  \bibinfo{volume}{56} (\bibinfo{year}{1995}) \bibinfo{pages}{445}.

\bibitem[{del Puerto et~al.(2008)del Puerto, Tiago, and Chelikowsky}]{Puerto08}
\bibinfo{author}{M.~L. del Puerto}, \bibinfo{author}{M.~L. Tiago},
  \bibinfo{author}{J.~R. Chelikowsky}, \bibinfo{journal}{Phys. Rev. B}
  \bibinfo{volume}{77} (\bibinfo{year}{2008}) \bibinfo{pages}{045404}.

\bibitem[{Gygi and Baldereschi(1989)}]{Gygi89}
\bibinfo{author}{F.~Gygi}, \bibinfo{author}{A.~Baldereschi},
  \bibinfo{journal}{Phys. Rev. Lett.} \bibinfo{volume}{62}
  (\bibinfo{year}{1989}) \bibinfo{pages}{2160}.

\bibitem[{Bechstedt et~al.(1992)Bechstedt, Sole, Cappellini, and
  L.Reining}]{Bechstedt92}
\bibinfo{author}{F.~Bechstedt}, \bibinfo{author}{R.~D. Sole},
  \bibinfo{author}{G.~Cappellini}, \bibinfo{author}{L.Reining},
  \bibinfo{journal}{Solid State Commun.} \bibinfo{volume}{84}
  (\bibinfo{year}{1992}) \bibinfo{pages}{765}.

\bibitem[{Bylander and Kleinman(1990)}]{Bylander90}
\bibinfo{author}{D.~M. Bylander}, \bibinfo{author}{L.~Kleinman},
  \bibinfo{journal}{Phys. Rev. B} \bibinfo{volume}{41} (\bibinfo{year}{1990})
  \bibinfo{pages}{7868}.

\bibitem[{Seidl et~al.(1996)Seidl, Gorling, Vogl, Majewski, and
  Seidl}]{Seidl96}
\bibinfo{author}{A.~Seidl}, \bibinfo{author}{A.~Gorling},
  \bibinfo{author}{P.~Vogl}, \bibinfo{author}{J.~A. Majewski},
  \bibinfo{author}{M.~Seidl}, \bibinfo{journal}{Phys. Rev. B}
  \bibinfo{volume}{53} (\bibinfo{year}{1996}) \bibinfo{pages}{3764}.

\bibitem[{Furthmuller et~al.(2002)Furthmuller, Cappellini, Weissker, and
  Bechstedt}]{Furthmuller02}
\bibinfo{author}{J.~Furthmuller}, \bibinfo{author}{G.~Cappellini},
  \bibinfo{author}{H.~C. Weissker}, \bibinfo{author}{F.~Bechstedt},
  \bibinfo{journal}{Phys. Rev. B} \bibinfo{volume}{66} (\bibinfo{year}{2002})
  \bibinfo{pages}{045110}.

\bibitem[{Lee and Wang(2006)}]{Lee06}
\bibinfo{author}{B.~Lee}, \bibinfo{author}{L.~W. Wang}, \bibinfo{journal}{Phys.
  Rev. B} \bibinfo{volume}{73} (\bibinfo{year}{2006}) \bibinfo{pages}{153309}.

\bibitem[{Fiorentini and Baldereschi(1995)}]{Fiorentini95}
\bibinfo{author}{V.~Fiorentini}, \bibinfo{author}{A.~Baldereschi},
  \bibinfo{journal}{Phys. Rev. B} \bibinfo{volume}{51} (\bibinfo{year}{1995})
  \bibinfo{pages}{17196}.

\bibitem[{Kersch and Fischer(2008)}]{Kersch08}
\bibinfo{author}{A.~Kersch}, \bibinfo{author}{D.~Fischer}, \bibinfo{journal}{J.
  Appl. Phys.} \bibinfo{volume}{106} (\bibinfo{year}{2008})
  \bibinfo{pages}{014105}.

\bibitem[{Gruning et~al.(2006)Gruning, Marini, and Rubio}]{Gruning06}
\bibinfo{author}{M.~Gruning}, \bibinfo{author}{A.~Marini},
  \bibinfo{author}{A.~Rubio}, \bibinfo{journal}{Phys. Rev. B}
  \bibinfo{volume}{74} (\bibinfo{year}{2006}) \bibinfo{pages}{161103(R)}.

\bibitem[{Harrison(1985)}]{Harrison85}
\bibinfo{author}{W.~A. Harrison}, \bibinfo{journal}{Phys. Rev. B}
  \bibinfo{volume}{31} (\bibinfo{year}{1985}) \bibinfo{pages}{2121}.

\bibitem[{Tran and Blaha(2009)}]{Tran09}
\bibinfo{author}{F.~Tran}, \bibinfo{author}{P.~Blaha}, \bibinfo{journal}{Phys.
  Rev. Lett.} \bibinfo{volume}{102} (\bibinfo{year}{2009})
  \bibinfo{pages}{226401}.

\bibitem[{Kresse and Furthm\"{u}ller(1996)}]{Kresse96}
\bibinfo{author}{G.~Kresse}, \bibinfo{author}{J.~Furthm\"{u}ller},
  \bibinfo{journal}{Phys. Rev. B} \bibinfo{volume}{54} (\bibinfo{year}{1996})
  \bibinfo{pages}{11169}.

\bibitem[{Kresse and Joubert(1999)}]{Kresse99}
\bibinfo{author}{G.~Kresse}, \bibinfo{author}{D.~Joubert},
  \bibinfo{journal}{Phys. Rev. B} \bibinfo{volume}{59} (\bibinfo{year}{1999})
  \bibinfo{pages}{1758}.

\bibitem[{Andersen and Jepsen(1984)}]{Andersen84}
\bibinfo{author}{O.~K. Andersen}, \bibinfo{author}{O.~Jepsen},
  \bibinfo{journal}{Phys. Rev. Lett.} \bibinfo{volume}{53}
  (\bibinfo{year}{1984}) \bibinfo{pages}{2571}.

\bibitem[{Andersen et~al.(1986)Andersen, Pawlowska, and Jepsen}]{Andersen86}
\bibinfo{author}{O.~K. Andersen}, \bibinfo{author}{Z.~Pawlowska},
  \bibinfo{author}{O.~Jepsen}, \bibinfo{journal}{Phys. Rev. B}
  \bibinfo{volume}{34} (\bibinfo{year}{1986}) \bibinfo{pages}{5253}.

\bibitem[{Turek et~al.(1997)Turek, Drchal, Kudrnovsky, Sob, and
  Weinberger}]{Turek97}
\bibinfo{author}{I.~Turek}, \bibinfo{author}{V.~Drchal},
  \bibinfo{author}{J.~Kudrnovsky}, \bibinfo{author}{M.~Sob},
  \bibinfo{author}{P.~Weinberger}, \bibinfo{title}{Electronic Structure of
  Disordered Alloy, Surfaces, and Interfaces}, \bibinfo{publisher}{Kluwer
  Academic Publishers}, \bibinfo{address}{Boston, MA}, \bibinfo{year}{1997}.

\bibitem[{Glotzel et~al.(1980)Glotzel, Segal, and Andersen}]{Glotzel80}
\bibinfo{author}{D.~Glotzel}, \bibinfo{author}{B.~Segal},
  \bibinfo{author}{O.~K. Andersen}, \bibinfo{journal}{Solid State Commun.}
  \bibinfo{volume}{36} (\bibinfo{year}{1980}) \bibinfo{pages}{403}.

\bibitem[{Adler(1962)}]{Adler62}
\bibinfo{author}{S.~L. Adler}, \bibinfo{journal}{Phys. Rev.}
  \bibinfo{volume}{126} (\bibinfo{year}{1962}) \bibinfo{pages}{413}.

\bibitem[{Wiser(1963)}]{Wiser63}
\bibinfo{author}{N.~Wiser}, \bibinfo{journal}{Phys. Rev.} \bibinfo{volume}{129}
  (\bibinfo{year}{1963}) \bibinfo{pages}{62}.

\bibitem[{Baroni and Resta(1986)}]{Baroni86}
\bibinfo{author}{S.~Baroni}, \bibinfo{author}{R.~Resta},
  \bibinfo{journal}{Phys. Rev. B} \bibinfo{volume}{33} (\bibinfo{year}{1986})
  \bibinfo{pages}{7017}.

\bibitem[{Skriver(1984)}]{Skriver84}
\bibinfo{author}{H.~L. Skriver}, \bibinfo{title}{The LMTO Method},
  \bibinfo{publisher}{Springer-Verlag}, \bibinfo{address}{Berlin},
  \bibinfo{year}{1984}.

\bibitem[{Vegard(1921)}]{Vegard21}
\bibinfo{author}{L.~Vegard}, \bibinfo{journal}{Z. Phys.} \bibinfo{volume}{5}
  (\bibinfo{year}{1921}) \bibinfo{pages}{17}.

\bibitem[{Monkhorst and Pack(1976)}]{Monkhorst76}
\bibinfo{author}{H.~J. Monkhorst}, \bibinfo{author}{J.~D. Pack},
  \bibinfo{journal}{Phys. Rev. B} \bibinfo{volume}{13} (\bibinfo{year}{1976})
  \bibinfo{pages}{5188}.

\bibitem[{Kootstra et~al.(2000)Kootstra, de~Boeij, and Snijders}]{Kootstra00}
\bibinfo{author}{F.~Kootstra}, \bibinfo{author}{P.~L. de~Boeij},
  \bibinfo{author}{J.~G. Snijders}, \bibinfo{journal}{Phys. Rev. B}
  \bibinfo{volume}{62} (\bibinfo{year}{2000}) \bibinfo{pages}{7071}.

\bibitem[{Ebina et~al.(1974)Ebina, Sato, and Takahashi}]{Ebina74}
\bibinfo{author}{A.~Ebina}, \bibinfo{author}{Y.~Sato},
  \bibinfo{author}{T.~Takahashi}, \bibinfo{journal}{Phys.Rev. Lett}
  \bibinfo{volume}{32} (\bibinfo{year}{1974}) \bibinfo{pages}{1366}.

\bibitem[{Brasil et~al.(1991)Brasil, Nahory, Turco-Sandorf, Gilchrist, and
  Martin}]{Brasil91}
\bibinfo{author}{M.~S. J.~P. Brasil}, \bibinfo{author}{R.~E. Nahory},
  \bibinfo{author}{F.~S. Turco-Sandorf}, \bibinfo{author}{H.~L. Gilchrist},
  \bibinfo{author}{R.~J. Martin}, \bibinfo{journal}{Appl. Phys. Lett.}
  \bibinfo{volume}{58} (\bibinfo{year}{1991}) \bibinfo{pages}{2509}.

\bibitem[{Wu et~al.(2003{\natexlab{b}})Wu, Walukiewicz, Yu, Shan, Ager, Haller,
  Miotkowski, Ramdas, and Su}]{Wu03b}
\bibinfo{author}{J.~Wu}, \bibinfo{author}{W.~Walukiewicz},
  \bibinfo{author}{K.~M. Yu}, \bibinfo{author}{W.~Shan}, \bibinfo{author}{J.~W.
  Ager}, \bibinfo{author}{E.~E. Haller}, \bibinfo{author}{I.~Miotkowski},
  \bibinfo{author}{A.~K. Ramdas}, \bibinfo{author}{C.~H. Su},
  \bibinfo{journal}{Phys. Rev. B} \bibinfo{volume}{68}
  (\bibinfo{year}{2003}{\natexlab{b}}) \bibinfo{pages}{033206}.

\bibitem[{Lin et~al.(2008)Lin, Chou, Fan, Ku, Ke, Wang, Yang, Chen, Chang, and
  Chia}]{Lin08}
\bibinfo{author}{Y.~C. Lin}, \bibinfo{author}{W.~C. Chou},
  \bibinfo{author}{W.~C. Fan}, \bibinfo{author}{J.~T. Ku},
  \bibinfo{author}{F.~K. Ke}, \bibinfo{author}{W.~J. Wang},
  \bibinfo{author}{S.~L. Yang}, \bibinfo{author}{W.~K. Chen},
  \bibinfo{author}{W.~H. Chang}, \bibinfo{author}{C.~H. Chia},
  \bibinfo{journal}{Appl. Phys.Lett.} \bibinfo{volume}{93}
  (\bibinfo{year}{2008}) \bibinfo{pages}{241909}.

\end{thebibliography}

%\end{thebibliography}
\end{document}